# Hedonic Models Incorporating ESG Factors for Time Series of Average Annual Home Prices

Jason R. Bailey*, W. Brent Lindquist, and Svetlozar T. Rachev

Department of Mathematics and Statistics, Texas Tech University, Lubbock, TX 79409-1042, USA

*Corresponding author: Jason.R.Bailey@ttu.edu

**Abstract**
Using data from 2000 through 2022, we analyze the predictive capability of the annual numbers of new home constructions and four available environmental, social, and governance factors on the average annual price of homes sold in eight major U.S. cities. We contrast the predictive capability of a P-spline generalized additive model (GAM) against a strictly linear version of the commonly used generalized linear model (GLM). As the data for the annual price and predictor variables constitute non-stationary time series, to avoid spurious correlations in the analysis we transform each time series appropriately to produce stationary series for use in the GAM and GLM models. While arithmetic returns or first differences are adequate transformations for the predictor variables, for the average price response variable we utilize the series of innovations obtained from AR($q$)-ARCH(1) fits. Based on the GAM results, we find that the influence of ESG factors varies markedly by city, reflecting geographic diversity. Notably, the presence of air conditioning emerges as a strong factor. Despite limitations on the length of available time series, this study represents a pivotal step toward integrating ESG considerations into predictive real estate models.

**Keywords**   real estate prices, environmental, social, and governance factors, generalized additive models, generalized linear models, stationarity in regression models

# 1. Introduction

Hedonic models are employed to analyze and predict average real estate prices via intrinsic and extrinsic factors. The average home price in a city plays an important role in the calculations made by potential homebuyers, particularly for low- and fixed-income buyers. Undoubtedly, the impacts of climate change and extreme weather will also affect the decisions made by potential homebuyers (as well as current homeowners) as the century progresses. Much work has been done in quantifying and modeling residence-based (e.g., lot area and number of bedrooms) and neighborhood-based (e.g., school zoning and homeowners' association fees) factors. The impact on real estate prices of the more recent developments in environmental, social, and governance (ESG) policies and factors have not been as well-analyzed. This paper contributes to that analysis.

We begin by briefly describing work that has been done to assess the impact of ESG factors on homebuilding and consumer decision-making. The sustainability factors of a property represent the environmental, social, and governance components of ESG. Resiliency to global warming, the risk of a natural disaster, and the installation of renewable energy systems are examples of environmental factors. Noise pollution, construction worker labor standards, and homeowner satisfaction are examples of social factors. Legal issues related to property owner practices, regulatory compliance with standards set at all governmental levels, and overall transparency are examples of governance factors.

Ma et. al. (2019) analyzed the impact of governmental policymaking processes on residential green energy additions and constructions. Even with a consumer base open to adopting environmentally friendly technologies, the cost bases, measured relative to non-green energy prices, play a large role in the adoption of such technologies. In particular, governmental policies on residential green energy subsidies that are too stringent can have an adverse effect on household installations.

Lauper et. al. (2013) analyzed the green home acquisition and installation process from the point of view of a homebuilder. Social factors (e.g., behavioral control and social norms) have meaningful impacts on the energy-relevant decisions made in homebuilding. Social policies and norms, such as low energy-consuming building certificates and awareness of available green technologies, have been shown to heighten consumer interest and spending on environmentally friendly appliances.

In addition to qualitative analyses based on consumer behavior, quantitative indices have been developed to provide guides to consumers in assessing home prices. Environmental factors (e.g., average maximum temperatures and flood risk) can be expected to play a role in homebuyer decisions (and, therefore, real estate pricing). Mahanama et. al. (2021) developed a natural disasters index to assess the level of future systemic risk caused by natural disasters. Their index used decades of property losses from the NOAA Storm Data to assess the main contributors to property losses. Although a homeowner's thought process can be very subjective, a quantification of the risk of extreme weather events represents an important step in translating subjective thought processes into quantitative factors for use in modeling. Indeed, a survey of research at the intersection of climate risks, housing, and mortgage markets was conducted, and natural disasters are expected to continue to weigh heavily on home prices (Contat et. al., 2023). Specifically, the risks of flooding and wildfires have been shown to correlate inversely with home prices, as higher risks of floods and wildfires result in discounts on home prices.

Intrinsic and extrinsic factors, including some ESG factors, have been used to describe the variance in (the logarithm of) the expected sales price of homes (Bailey et. al., 2022). When ESG factors (accessible by the elderly and disabled, presence of central air conditioning, "green home" rating, and waterfront location) commonly available on real estate vendor websites were included, minor improvements were observed in the model adjusted $R^2$ values. Although the model results were city-dependent, the potential impact that such ESG factors had in assessing home prices was established.

Other ESG factors not currently featured on real estate vendor sites, such as the impacts of air pollution, have been explored. For example, an analysis of the closure of a toxic site leading to changes in atmospheric pollution levels found that the corresponding drop in $SO_2$ levels correlated with an average house price increase of 6% (Lavaine, 2019). However, the average price of flats decreased by 9%, suggesting that the impacts of refinery closures and changes in air pollution levels have heterogeneous effects on subsamples.

The usual application of a hedonic home-pricing model is "cross-sectional", consisting of a data set of response (price) and predictor (e.g., number of bedrooms, bathrooms, home size, etc.) variables for a sample of homes. Implicit in the cross-sectional analysis is the assumption that the data set represents independent, identically distributed, random samples reflective of the pricing structure in a particular geographic area. In contrast, our application in this paper is to time series data. For a geographical area, specifically a city, the data set (Section 2.1) consists of the average annual home price, number of homes sold per year, and yearly values for each of four available ESG factors. As we show (Appendix A), each time series is non-stationary (exhibiting strong year-by-year trend), which can produce spurious correlation in fits by hedonic models.

This paper pursues three goals. The first is the determination of an appropriate transformation into stationary form for each time-series of annual data. The second is to evaluate the effectiveness and accuracy of the application, to these transformed series, of a P-spline-based generalized additive model (GAM) compared to a generalized linear model (GLM). The analysis used data from eight cities, which were chosen to represent variations in geography (the Sun and Frost Belts, the Pacific and Atlantic coasts, and "Middle America"), primary economic activity, and population size and density. As each transformed time-series is "de-trended", it represents values from a fixed-mean and fixed-variance random variable. Using principal component analysis, our third objective is to investigate the time series for the residual term, aggregated across cities, from each of the GAM and GLM fits to determine the presence of further latent random variables. As our model is deliberately parsimonious in terms of the number of predictor factors, we hypothesize the presence of such latent variables.

## 2. Materials and Methods

*2.1 Price and Factor Data*

Price and factor data were acquired from Zillow[1]. Our data set is composed of completed sale transactions of homes[2] each year for the years 2000 through 2022 for eight cities[3]. For each year and city, the data set consists of the average home sale price (Av Price), the total number of homes constructed (New Homes), and four ESG factors: the number of homes with central air conditioning (Central AC), the number of green-rated homes (Green), the number of homes that are considered accessible to the elderly and disabled (Accessible), and the number of homes along a waterfront (Waterfront)[4]. The eight cities studied were Atlanta, GA (ATL), Austin, TX (AUS), Columbus, OH (COL), Jacksonville, FL (JAX), Nashville, TN (NAS), Oklahoma City, OK (OKC), Portland, OR (POR), and Seattle, WA (SEA). Table A2 in Appendix A summarizes the full data set for ATL.

*2.2 Generalized Additive and Linear Models*

A GAM model relates a univariate response variable $Y_t$ to a set of predictor variables (factors) $x_{k,t}$, $k = 1, \ldots, m$. (Here, the subscript $t = 1, \ldots, \tau$ indicates the observed set of values of the response and predictor variables. In our application, $t$ indicates yearly time values.) Specifically, the GAM model relates the expected value $\mu_t = E[Y_t]$ to the predictor values via

$$g(\mu_t) = \beta_0 + f_1(x_{1,t}) + f_2(x_{2,t}) + \cdots + f_m(x_{m,t}), \quad t = 1, \ldots, \tau. \tag{1}$$

The model assumes $Y_t \sim EF(\mu_t, \theta)$, where $EF(\mu_t, \theta)$ denotes the exponential family of distributions having mean $\mu_t$ and scale parameter $\theta$. Choice of the link function $g(\cdot)$ relates expected values of the average $\mu_t$ to the factors via

$$\mu_t = g^{-1}\left(\beta_0 + f_1(x_{1.t}) + f_2(x_{2,t}) + \cdots + f_m(x_{m,t})\right) + \varepsilon_t, \tag{2}$$

where $\varepsilon_t$ denotes the residual error that is not captured by the model. The identity function was used for $g(\cdot)$, and P-splines (Eilers and Marx, 1996) were used for the functions $f_j(\cdot)$. Such P-splines minimize the penalized sum of squares,

$$\sum_{i=1}^{\tau}\left(Y_\tau - \sum_{j=1}^{m} f_j(x_{j,t})\right)^2 + \sum_{j=1}^{m} \lambda_j \int f_j''(z)^2 dz, \tag{3}$$

where the tuning parameters $\lambda_j > 0$ determine the weight assigned to the smoothness of each function. The values $x_{tj}$ are referred to as the knots for the function $f_j(\cdot)$.

---

[1] Data from https://www.zillow.com/homes/ was collected by specifying the city in the search field and then the entries for all filters as provided in Table A1 in Appendix A.
[2] Home types considered are specified in the appropriate filter in Table A1.
[3] Note that the data applies only to homes constructed within the city boundaries and not to homes within the associated Metropolitan Statistical Area.
[4] Specifically, three of the factors are environmental and one (accessibility) is social, although all four are often influenced by local policies.

The results acquired from this GAM were compared to those from a standard GLM of the form,

$$g(E_Y(Y|X)) = \beta_0 + \beta_1 x_1 + \cdots + \beta_m x_m + \xi \equiv X\beta + \xi. \tag{4}$$

In (4), matrix notation is used to represent the response and predictor values: $Y = [Y_1, \ldots, Y_\tau]^T$ is the column vector of response values, $\beta = [\beta_0, \beta_1, \ldots, \beta_m]^T$ is the column vector of unknown parameters, $\xi = [\varepsilon_0, \varepsilon_1, \ldots, \varepsilon_\tau]^T$ is the column vector of residuals, and $X$ is a $\tau \times (m+1)$ matrix. The first column of $X$ is a vector of ones, while column $j$, $j = 2, \ldots, m+1$, is the vector $x_j = [x_{j,1}, \ldots, x_{j,\tau}]^T$ of values for factor $x_j$. As in the GAM model, we used the identity function for $g(\cdot)$, which reduced (4) to a pure linear model.

In the GAM model (1), use of a non-linear function $f_j(\cdot)$ provides non-linear dependence on the time-dependent value of its argument $x_{j,t}$. In the GLM model (4), the corresponding coefficient $\beta_j$ is constant, which results in linear dependence on the time-dependent value of $x_{j,t}$. Thus, a non-linear form for the $f_j(\cdot)$ enables a greater fitting accuracy for the GAM compared to the GLM. On the other hand, the GLM provides a superior model for prediction. The form of the function $f_j(\cdot)$ depends on the known values of its knots. Accurate prediction by GAM requires knowledge of future knot values to a much greater degree than does the GLM[5].

*2.3 Transformation to Stationary Time Series*

We wish to compare the accuracy of models (1) and (4) in predicting the expected value $\mu_t = E[Y_t]$ of average annual home price $Y_t$ in terms of the five factors $x_j, j = 1, \ldots, 5$, described in section 2.1 for each of the eight cities. The data set consists of 23 years of price and factor values. However, to avoid spurious correlations in the factor analyses, it is necessary that the time series for the response variable and each factor be stationary. Stationarity of each time series was investigated via the augmented Dickey-Fuller (ADF) test by using the random walk model[6]

$$\Delta y_t = \alpha y_{t-1} + \sum_{i=1}^{q} \delta_i \Delta y_{t-i} + \varepsilon_t. \tag{5}$$

The statistic $\text{DF}_\alpha = \hat{\alpha}/\text{SE}(\hat{\alpha})$ [7] is used to test the hypotheses $H_0: \alpha = 0$ (the existence of a unit root) against $H_A: \alpha < 0$. The rejection of the null hypothesis through a sufficiently small $p$-value suggests that no unit root is present, and that stationarity can be assumed.

As illustrated in Fig. A1 (Appendix A) for ATL and as verified by the ADF test, stationarity could not be inferred (at any reasonable level of statistical significance) for any factor or price time series for any of the eight cities. As discussed in Appendix A and illustrated for ATL in Fig. A2, use of the arithmetic return series for each predictor factor produced transformed time series that were acceptable. There were three exceptions for which the arithmetic return time series had to be replaced by simple first differences to avoid division by zero: the Accessible factor for Seattle, the Green factor for Columbus, and the Waterfront factor for three cities. As the first-difference time series for the Waterfront factor for all eight cities had very acceptable $p$-values (below 1.5%), for consistency we used the first-difference time series for the Waterfront factor for all cities. Thus, the predictor variables used in (1) and (4) represent transformed series (with the transformation being either arithmetic returns or first differences). The specific transformation used for each factor is summarized in Table A4 in Appendix A.

---

[5] Restated in the context of the P-splined-based GAM and the GLM used here, extrapolation using polynomials is much less accurate than extrapolation using a linear least-squares fit.
[6] In (5), we use a generic notation $y_t$ to denote the time series being tested.
[7] SE(·) denotes standard error.

Table 1 provides the $p$-values for the ADF tests computed on each of the transformed predictive factor series. The transformed time series for each factor is (assumed) stationary at a 5% significance level, with four exceptions: New Homes and Central AC for ATL and Accessible for JAX and OKC. Only one (Accessible for JAX) is not significant at the 10% level.

**Table 1.** ADF test $p$-values for each transformed time series by city.

| Factor | ATL | AUS | COL | JAX | NAS | OKC | POR | SEA |
|---|---|---|---|---|---|---|---|---|
| New Homes | 0.074 | ** | 0.017 | 0.034 | ** | 0.021 | ** | 0.014 |
| Central AC | 0.069 | 0.015 | 0.013 | 0.032 | 0.012 | 0.036 | ** | ** |
| Green | ** | ** | ** | ** | ** | ** | ** | ** |
| Accessible | 0.012 | ** | ** | 0.244 | ** | 0.076 | ** | ** |
| Waterfront | 0.015 | ** | ** | ** | ** | ** | ** | ** |
| Av Price Innovations | 0.020 | ** | 0.036 | 0.053 | 0.095 | 0.055 | 0.020 | 0.097 |

\*\* Indicates $p$-value $< 0.01$.

Neither arithmetic returns nor first (nor second) differences were sufficient to achieve stationarity for the average price time series. To obtain stationarity, we resorted[8] to fitting an AR($q$)-ARCH(1)-Student's-$t$ model,

$$r_t - \mu_r = \sum_{i=1}^{q} \varphi_i(r_{t-i} - \mu_r) + \epsilon_t,$$
$$\epsilon_t = \sigma_t z_t, \quad z_t \sim t_\nu,$$
$$\sigma_t^2 = \omega + \alpha_1 \epsilon_{t-1}^2,$$
(6)

to the arithmetic return series $r_t = (Y_t - Y_{t-1})/Y_{t-1}$ of the average price. In (6), $t_\nu$ denotes the Student's-$t$ distribution with $\nu$ degrees of freedom. A fit was judged satisfactory if the innovation series $z_t$ was determined to be stationary (as verified by the ADF test). For each city, we chose the smallest value of $q$ that produced a stationary innovation series. As detailed in Table A4 in Appendix A, the value $q = 1$ was sufficient for four cities, while $q = 2$ was required for the remaining four. The $p$-values obtained for the price innovation time series are also listed in Table 1. Four are significant at the 5% level, and the remaining are significant at the 10% level. The resulting innovation time series, $z_t$, replaced the average price series in the GAM and GLM fits.

---

[8] We tested a variety of ARFIMA-GARCH models before settling on AR($q$)-ARCH(1) with $q = \{1, 2\}$. We desired an ARFIMA-GARCH model which was as parsimonious as possible in the number of coefficients to be fit.

*2.4 Principal Component Analysis for Additional Systematic Factors*

As noted in Section 2.3, the GAM and GLM models were applied to response variables consisting of "average-price innovation" time series and either arithmetic return or first-differenced transformed predictor variable time series. As a result of the transformations, each time series is reduced to 22 (rather than 23) years of observations (2001 through 2022). For each city, the difference between the AR($q$)-ARCH(1) derived innovation and the regression model fit results in 22 residual values (one per year). These residuals can be assembled in a matrix $R = \{\varepsilon_{t,k}\}$, $t = 1, \dots, 22$, $k = 1, \dots, 8$ (i.e., the city index[9] ). To determine whether systemic factors remained in the residuals, we performed a principal component analysis by computing the eigenvalues and eigenvectors of the variance-covariance matrix $R^T R$ (Rachev et. al., 2007). The eigenvectors correspond to the principal components (ordered, as usual, so that the first principal component has the largest magnitude eigenvalue, etc.). We refer to extreme value theory (de Haan and Ferreira, 2006) to analyze the type of decay exhibited by the explained variances[10] associated with the principal components. Consider the decaying discrete exponential distribution,

$$f_1(x) = \frac{1}{\beta}(1-\beta)^{(x-1)} = \frac{1}{\beta(1-\beta)} e^{x \ln(1-\beta)}, \quad x = 0, 1, \dots, \quad \beta \in (0, 1), \tag{7}$$

and the decaying power-law zeta distribution,

$$f_2(x) = \frac{1}{\zeta(b)} x^{-b}, \quad x = 1, 2, \dots, \quad b > 1, \tag{8}$$

where $\zeta(b)$ is the Riemann zeta function and $x$ is the index of the principal component. The relative changes with respect to $x$ in these two distributions are

$$R_1(x) = \frac{f_1(x+1) - f_1(x)}{f_1(x)} = -\beta, \quad R_2(x) = \frac{f_2(x+1) - f_2(x)}{f_2(x)} = \left(\frac{x}{1+x}\right)^b - 1. \tag{9}$$

As the magnitude of $R_1(x)$ is independent of $x$, each component of the exponential fit has the same relative drop in importance. On the other hand, for any $b > 1$, the magnitude of $R_2(x)$ decreases as higher components are added; thus, additional components add less value to the model. Power decay suggests that noise dominates the residuals, whereas exponential decay suggests that systemic factors continue to be unaccounted for (de Haan and Ferreira, 2006). If $f(x)$ represents the observed distribution of proportion of variance, plots of $\ln f(x)$ vs. $x$ compared with $\ln f(x)$ vs. $\ln(x)$ will distinguish between exponential and power-law tail behavior.

---

[9] We index the cities in alphabetical order.
[10] The explained variance associated with each principal component is the ratio of its eigenvalue to the sum of all eigenvalues.

# 3. Results

## 3.1 GLM and GAM Results

The GLM and GAM models were run using the R *lm* (linear model) function and the *gam* package (Hastie, 2023). Table 2 displays the $p$-values associated with the various factors for each city as fit by the GLM and GAM. Note that the $p$-values for ATL are identical under both the GAM and the GLM. For this city, the GAM P-splines simplified to linear terms and became identical to the GLM.

Table 2. Significance ($p$-value) of the factors in the GLM and GAM fits.

| Factor | ATL | AUS | COL | JAX | NAS | OKC | POR | SEA |
|---|---|---|---|---|---|---|---|---|
| GLM | | | | | | | | |
| New Homes | 0.747 | 0.189 | 0.184 | 0.103 | 0.025 | 0.515 | 0.176 | 0.632 |
| Accessible | 0.467 | 0.994 | 0.169 | 0.315 | 0.585 | ** | 0.353 | 0.320 |
| Central AC | 0.594 | 0.234 | 0.169 | 0.117 | 0.024 | 0.700 | 0.550 | 0.879 |
| Green | 0.500 | 0.100 | 0.249 | 0.633 | 0.247 | 0.116 | 0.191 | 0.457 |
| Waterfront | 0.629 | 0.975 | 0.838 | 0.807 | 0.041 | 0.929 | 0.855 | 0.242 |
| Adj. $R^2$ | −0.167 | −0.061 | 0.144 | 0.159 | 0.218 | 0.504 | −0.525 | 0.226 |
| GAM | | | | | | | | |
| New Homes | 0.747 | 0.151 | 0.100 | 0.017 | 0.091 | 0.152 | 0.017 | 0.555 |
| Accessible | 0.467 | 0.945 | 0.031 | 0.021 | 0.677 | ** | 0.720 | 0.169 |
| Central AC | 0.594 | 0.240 | 0.063 | 0.027 | 0.085 | 0.015 | 0.032 | 0.997 |
| Green | 0.500 | 0.019 | 0.356 | 0.188 | 0.102 | ** | 0.073 | 0.363 |
| Waterfront | 0.629 | 0.646 | 0.069 | 0.085 | ** | 0.984 | 0.123 | 0.462 |
| Adj. $R^2$ | −0.167 | 0.388 | 0.518 | 0.560 | 0.703 | 0.855 | 0.468 | 0.349 |

** Indicates $p$-value $< 0.01$.

Table 3. Summary of marginal significances in Table 2 using a $p$-value threshold of 10%.

| | Number of significant factors | | | | | | | |
|---|---|---|---|---|---|---|---|---|
| Model | ATL | AUS | COL | JAX | NAS | OKC | POR | SEA |
| GLM | 0 | 1 | 0 | 0 | 3 | 1 | 0 | 0 |
| GAM | 0 | 1 | 4 | 4 | 3 | 3 | 3 | 0 |

| | Number of cities for which a factor is significant | | | | |
|---|---|---|---|---|---|
| Model | New Homes | Accessible | Central AC | Green | Waterfront |
| GLM | 1 | 1 | 1 | 1 | 1 |
| GAM | 4 | 3 | 5 | 3 | 3 |

With only a small set of factors, we evaluate the significance of each based upon a level of 0.10. Table 3 summarizes the number of significant factors for each city as well as the number of cities for which each factor was found to be significant. In either marginal view (i.e., by city or by factor), the number of significant quantities under GAM equaled or exceeded that under GLM. Notable differences in the number of significant factors occurred for COL, JAX, OKC, and POR. Increases in the significant number occurred for all five factors, particularly for New Homes and Central AC.

Table 2 also presents the adjusted $R^2$ values obtained from the model fits. These values are reflective of the marginal significance numbers summarized in Table 3. Figure 1 presents a box-and-whisker summary of the spread of the adjusted $R^2$ values for each model. The non-linear GAM model produces consistently better values. The fact that some values are negative, particularly for GLM, indicates model inappropriateness. Given the small number of predictor variables, the large adjusted $R^2$ values for the GAM model were unexpected and indicate a direction for future investigation.

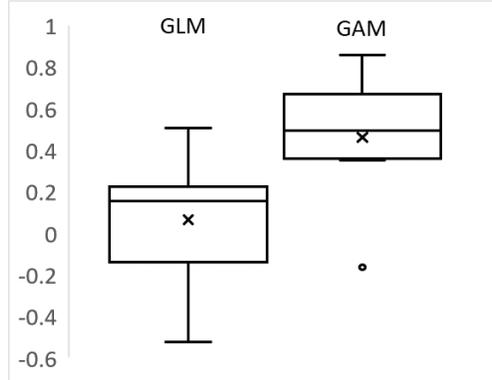

**Figure 1.** Box-and-whisker summary of the adjusted $R^2$ values of Table 2 for the GLM and GAM fits.

The results illustrate the potential for using ESG return (or first difference) factors in modeling average home price innovation time series for cities. The GAM results indicate that such relationships are nonlinear. The nonlinear nature of GAM is able to distinguish the predictive capability of the ESG factors, in particular the influence of central air conditioning, on average home prices. The results show conformity with the geographic locations and demographic profiles of the cities. As an example, consider the Waterfront factor. Water-body percentage (the percent of city area that is of a body of water) is a proxy (though not always an accurate one[11]) for waterfront acreage. The percentage of each city's area comprised of bodies of water is provided in Table 4.

**Table 4.** Percent of water area and percent of seniors living alone, by city

|            | ATL | AUS | COL | JAX  | NAS | OKC | POR | SEA  |
|------------|-----|-----|-----|------|-----|-----|-----|------|
| Water area | 0.7 | 2.0 | 2.6 | 14.5 | 4.2 | 2.3 | 7.9 | 40.9 |
| Seniors    | 3.8 | 4.6 | 7.2 | 7.9  | 8.2 | 13.4| 9.0 | 4.1  |

[1] source: 2023 US Census Gazetteer Files    [2] source: 2010 US Census

Consider the scatterplot of Waterfront GAM $p$-values versus water-body percentage show in Fig. 2. As the proxy is approximate, we look for a fuzzy relationship by dividing the plot into four quadrants. Significant occupancy in the (low, high) and (high, low) quadrants indicates a fuzzy inverse relationship between the water-body percentage and Waterfront $p$-value. Three cities (ATL, AUS, and OKC) occupy the (low, high) quadrant with three (COL, JAX, and NAS) occupying the (high, low) quadrant. Two cities (POR and SEA) occupy the (high, high) quadrant with POR lying very close to the (high, low) quadrant. SEA's water-body percentage is a poor proxy for waterfront area since a large fraction of the water-body area (Puget Sound and Lake Washington) is distant from the shoreline.

---

[11] If two cities border a body of water in the United States, then the common city boundary often divides the body of water along a line medial to the city shorelines. Thus, two or more cities bordering a large and contained body of water can have large water-body percentages but relatively short shorelines.

Similarly, we consider the percentage of seniors living alone as a proxy for the Accessible factor. Table 4 also shows the 2010 census results on the percentage of seniors living alone in each of the eight cities, and Fig. 2 shows the relevant scatter plot and quadrants. Again, three cities (ATL, AUS, and SEA) occupy the (low, high) quadrant with three occupying (COL, JAX, and OKC) the (high, low) quadrant, thus indicating a fuzzy inverse relationship established by six of the eight cities.

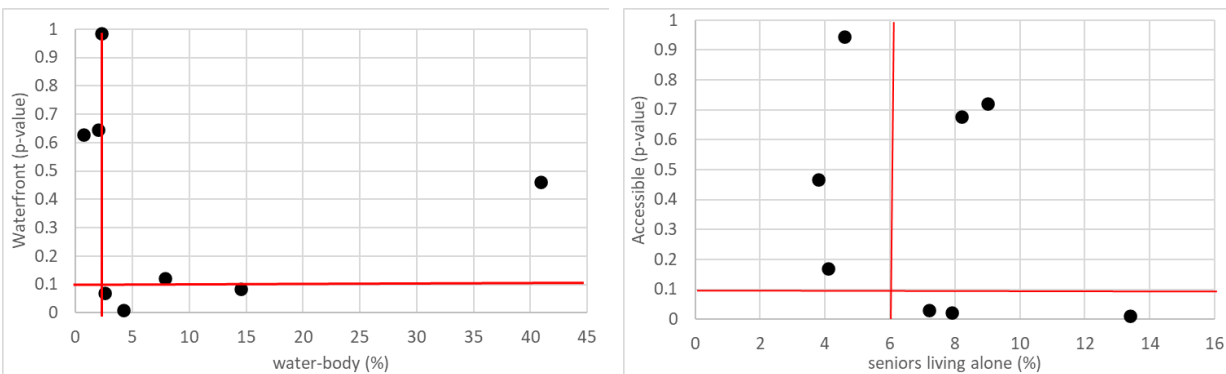

**Figure 2.** (left) Waterfront $p$-value versus water-body percentage and (right) Accessible $p$-value versus percentage of seniors living alone for the GAM fits.

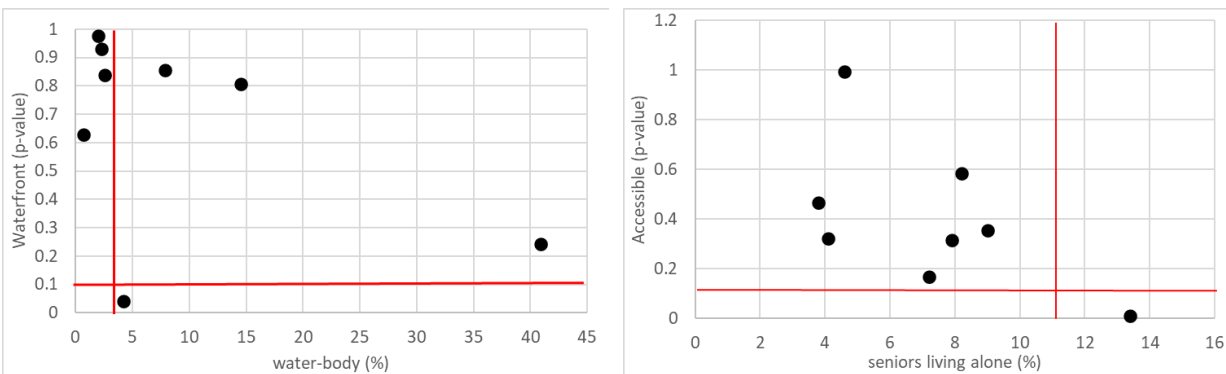

**Figure 3.** (left) Waterfront $p$-value versus water-body percentage and (right) Accessible $p$-value versus percentage of seniors living alone for the GLM fits.

For comparison purposes, Fig. 3 presents these quadrant plots for the GLM fits. Maintaining a $p$-value threshold of 0.01 leads to no meaningful inverse relationship between Waterfront $p$-value and water-body percentage. One might argue there is a fuzzy inverse relationship between percentage of seniors living along and Accessible $p$-value in the GLM results, but this would be based upon a (high, low) quadrant occupancy of a single city (OKC).

*3.2 Principal Component Analysis and Residuals Results*

Tables B1 and B2 in the Appendix provide the residual matrices $R$ for each model. Table 5 displays the proportion of variance obtained for each of the identified components for each model.

Fits of the exponential and power-law decays (7) and (8) to the proportions of variance data in Table 5 are presented in Fig. 4 along with the $R^2$ and mean squared error (MSE) results for each.

Table 5. Proportion of explained variance by principal component (PC)

| Model | PC 1 | PC 2 | PC 3 | PC 4 | PC 5 | PC 6 | PC 7 | PC 8 |
|---|---|---|---|---|---|---|---|---|
| GLM | 0.453 | 0.205 | 0.111 | 0.087 | 0.061 | 0.041 | 0.028 | 0.014 |
| GAM | 0.319 | 0.209 | 0.144 | 0.138 | 0.075 | 0.050 | 0.039 | 0.028 |

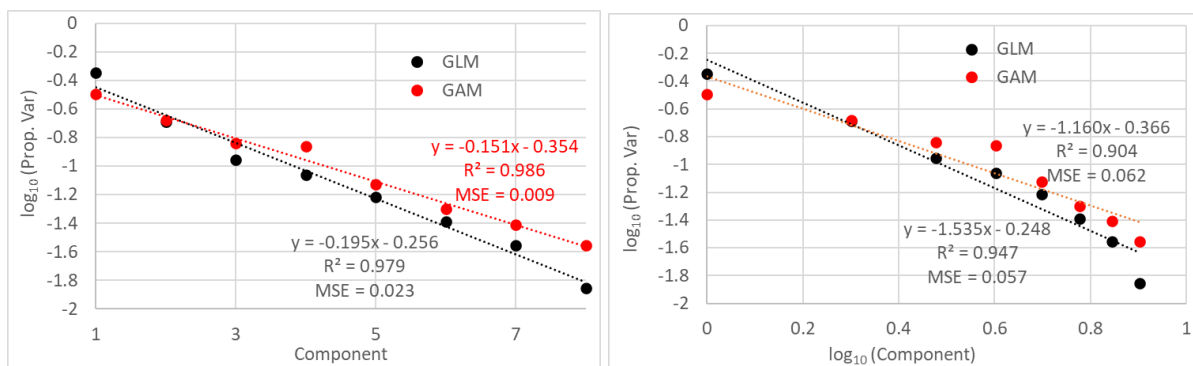

**Figure 4.** (left) Exponential and (right) power-law fits to the proportions of variance obtained for eight components arising from the principal component analysis for the GLM and GAM fits to all eight cities.

Visually and through the quantitative $R^2$ and MSE values, it is clear that the exponential fit does a better job of describing the data. Thus, we conclude that systemic factors not included in our models exist in the residuals. This is fully anticipated, as we had no expectation that a model for average annual home prices that considered only new home constructions and four ESG factors would encompass all significant factors. Moreover, a comparison of the $R^2$ and MSE for the GAM and GLM exponential fits further supports our findings from the adjusted-$R^2$ numbers and $p$-values that the GAM provides a model that is superior to the GLM.

## 4. Conclusion

Our results demonstrate that P-spline GAMs possess strong predictive capabilities for the expected value of the average annualized home sale price in major U.S. cities on ESG factors. The results stand in stark contrast to the GLMs. Although each factor in the GAM was significant for multiple cities, some factors (particularly central air conditioning) were especially prevalent. As climate change continues to warm the planet, cities at more northerly latitudes which would otherwise not experience hotter temperatures will experience higher rates of central air conditioning. Therefore, we would expect that the significance of available central air conditioning would increase as the century progresses for both average annualized real estate prices and individual home prices. Overall, the results of the eight surveyed cities strongly suggest that the significance of ESG factors is very city dependent.

One weakness of the current data set is the length of each time series. Yearly data points over 23 years, results in first difference, return and innovation time series of length 22, reducing the effective sample size needed for fits by the GAM, GLM and ARFIMA-GARCH-based models. It would be better to have access to monthly data over the same 23-year time period. Unfortunately, we had no access to such data for this study.

**Data Availability**. The study's data is available upon request to the corresponding author.
**Code Availability**. The study's source code is available upon request to the corresponding author.
**Conflicts of Interest.** The authors declare no conflict of interest.

**Appendix A**

**Table A1.** Filter values used for Zillow data

| Filter | Input | Filter | Input |
|---|---|---|---|
| Status | Sold | | |
| Price Range | MIN: $50k, MAX: $10M | | |
| Number of Bedrooms | 1+ | | |
| Number of Bathrooms | 1+ | | |
| Home Type | Houses, Townhomes, Multi-Family, and Condos/Co-ops | | |
| **More Filters** | | | |
| Max HOA | Any | Must have A/C | ESG[2] |
| Parking Spots | Any | Must have pool | NS |
| Square Feet | MIN: 500, MAX: NS[1] | Waterfront | ESG[2] |
| Lot Size | MIN: NS, MAX: NS | City | NS |
| Year Built | MIN: 2000, MAX: 2022 | Mountain | NS |
| Has basement | NS | Park | NS |
| Single-story only | ACC[3] | Water | NS |
| Hide 55+ communities | NS | Sold in Last | 36 months |
| **Keywords** | | | |
| "Green", "Green Home" | ESG | "Accessible" | ACC |

[1]NS = Not specified     [2] "Yes" when filtering for those houses and "NS" otherwise
[3]Single-story only and/or classified as "Accessible"

Table A2 provides the price and factor data for the city of Atlanta. Fig. A1 plots these 22-year time series. Visual inspection clearly shows trends (non-stationarity) in each time series. The *p*-values obtained from the ADF test run on each of these times series are provided in Table A3. All *p*-values are strongly indicative of non-stationary time series. A common method to transform a non-stationary time series $x_t$ into a stationary one is through first differences $\Delta x_t = x_t - x_{t-1}$ or, equivalently, arithmetic returns $r_t = \Delta x_t / x_{t-1}$.[12] Use of arithmetic returns is preferred as its use on different time series produces transformed series of comparable magnitudes. Fig. A2 plots the time series for Fig. A1 in terms of their arithmetic return. Visually, the trends are eliminated (or vastly reduced). Fig. A3 plots the return series for the price and factor series, while Table A3 presents the resultant *p*-values from the ADF test. All are vastly improved, indicating stationarity (at a threshold significance of 7.5%), with the exception of the price series. Fits of an AR(*q*)-ARCH(1)-Student's *t* model to the price series for ATL produced the best results for $q = 2$. The resultant innovation time series is shown in Fig. A3, and the *p*-value from the ADF test for the innovation time series is provided in Table A3.

**Table A2.** Price and factor data for ATL.

| Year | Av Price | New Homes | Accessible | Central AC | Green | Waterfront |
|---|---|---|---|---|---|---|
| 2000 | 174500 | 456 | 43 | 454 | 20 | 31 |
| 2001 | 187800 | 718 | 85 | 716 | 38 | 38 |
| 2002 | 196400 | 919 | 112 | 873 | 46 | 79 |
| 2003 | 203400 | 649 | 38 | 615 | 26 | 43 |
| 2004 | 211700 | 1267 | 139 | 1235 | 40 | 107 |
| 2005 | 222000 | 1842 | 140 | 1815 | 55 | 142 |
| 2006 | 229200 | 1775 | 171 | 1718 | 37 | 139 |
| 2007 | 233800 | 1451 | 124 | 1386 | 36 | 99 |
| 2008 | 225500 | 916 | 145 | 902 | 19 | 52 |
| 2009 | 212000 | 427 | 41 | 401 | 30 | 74 |
| 2010 | 195600 | 330 | 46 | 305 | 47 | 31 |
| 2011 | 180500 | 89 | 2 | 88 | 7 | 5 |
| 2012 | 172900 | 105 | 1 | 114 | 6 | 14 |
| 2013 | 183400 | 168 | 2 | 174 | 7 | 11 |
| 2014 | 200900 | 167 | 3 | 182 | 13 | 8 |
| 2015 | 216600 | 283 | 5 | 277 | 16 | 14 |
| 2016 | 232400 | 319 | 5 | 305 | 11 | 23 |
| 2017 | 249100 | 404 | 6 | 394 | 12 | 26 |
| 2018 | 269600 | 451 | 2 | 432 | 25 | 28 |
| 2019 | 286400 | 629 | 11 | 823 | 31 | 26 |
| 2020 | 303200 | 1225 | 31 | 968 | 32 | 75 |
| 2021 | 351300 | 1091 | 54 | 1019 | 55 | 88 |
| 2022 | 430000 | 740 | 29 | 760 | 51 | 67 |

---

[12] Higher-order differences may be required if the time series is integrated of an order higher than one.

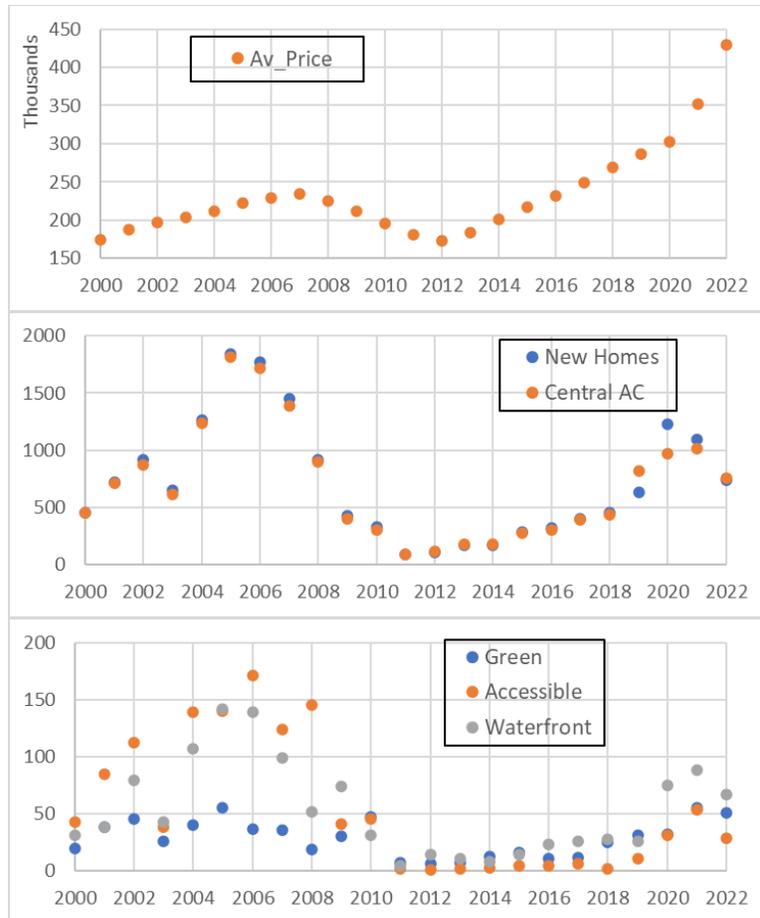

**Figure A1.** Time series for year-averaged home sale price (Av Price) and the factors New Homes, Accessible, Central AC, Green, and Waterfront for the city of Atlanta for the years 2000 through 2022. (Source: Zillow)

**Table A3.** Significance ($p$-value) of the time series for ATL

| Factor | New Homes | Accessible | Central AC | Green | Water-front | Price |
|---|---|---|---|---|---|---|
| raw data (Fig. A1) | 0.729 | 0.368 | 0.757 | 0.544 | 0.632 | >0.990 |
| arithmetic return | 0.074 | 0.012 | 0.069 | ** | ** | 0.943 |
| AR(2)-ARCH(1) innovation | na[1] | na | na | na | na | 0.020 |

** Indicates $p$-value $< 0.01$.    [1] Not applicable.

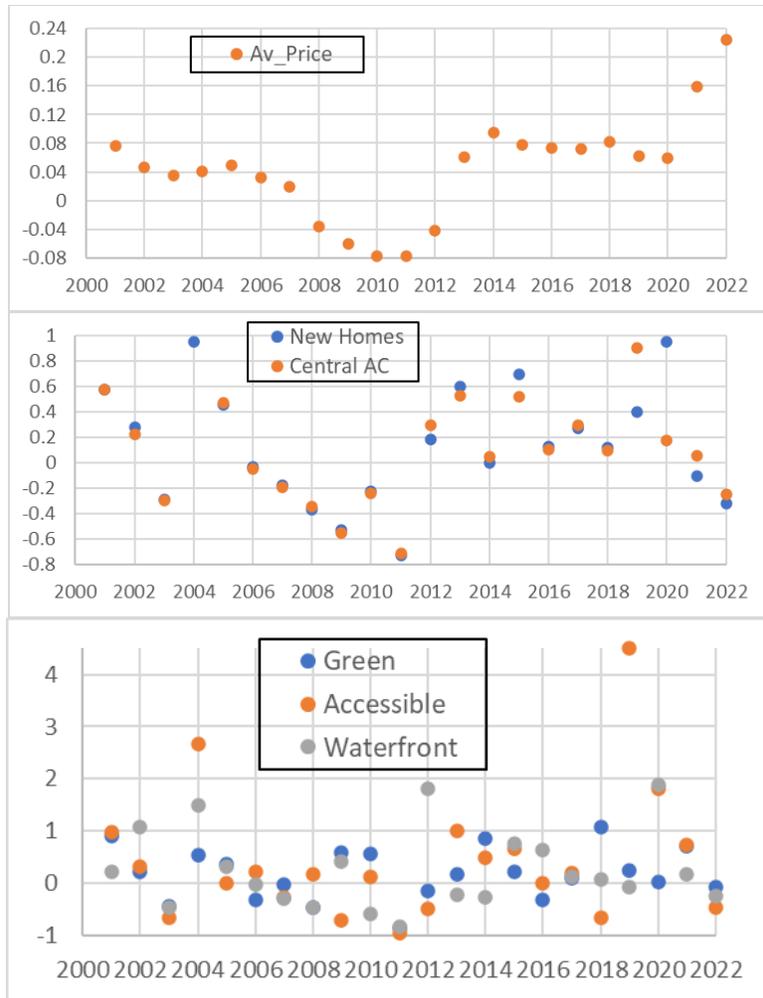

**Figure A2.** Arithmetic return series for the times series displayed in Fig. A1.

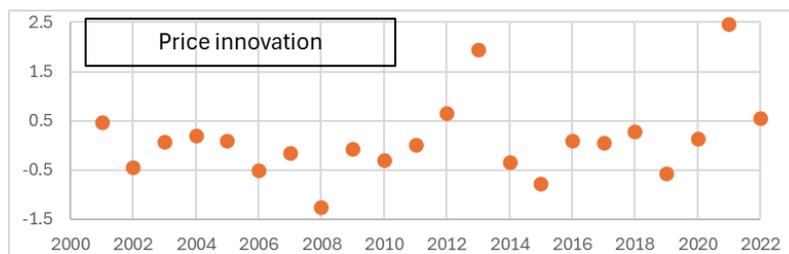

**Figure A3.** Price innovation time series for ATL.

Table A4 summarizes the transformations that were used on each time series, while Table 1 (Section 2.3) summarizes the *p*-values obtained from the ADF tests.

**Table A4.** Type of transformed series used in GAM and GLM regressions.

| Factor     | ATL      | AUS | COL | JAX | NAS | OKC | POR | SEA |
|------------|----------|-----|-----|-----|-----|-----|-----|-----|
| New Homes  | rtn[1]   | rtn | rtn | rtn | rtn | rtn | rtn | rtn |
| Central AC | rtn      | rtn | rtn | rtn | rtn | rtn | rtn | rtn |
| Green | rtn | rtn | fd[2] | rtn | rtn | rtn | rtn | rtn |
| Accessible | rtn | rtn | rtn | rtn | rtn | rtn | rtn | fd |
| Waterfront | fd | fd | fd | fd | fd | fd | fd | fd |
| Av Price Innovation | $q=2$[3] | $q=1$[4] | $q=1$ | $q=2$ | $q=1$ | $q=1$ | $q=2$ | $q=2$ |

[1] rtn: arithmetic return  [2] fd: first difference  [3] AR(2)-ARCH(1)  [4] AR(1)-ARCH(1)

## Appendix B

**Table B1.** Residual values ($US) for the GLM fit on each city by year.

| Year | ATL | AUS | COL | JAX | NAS | OKC | POR | SEA |
|---|---|---|---|---|---|---|---|---|
| 2001 | 0.068 | -0.229 | -0.082 | 0.050 | -0.101 | 0.001 | 0.218 | -0.025 |
| 2002 | -0.793 | -0.282 | -0.086 | 0.292 | -0.540 | -0.207 | -0.591 | -0.655 |
| 2003 | 0.246 | -0.608 | 0.394 | 0.244 | -0.332 | 0.253 | -0.100 | 0.773 |
| 2004 | -0.191 | 0.057 | -0.079 | 0.341 | -0.088 | -0.043 | 0.826 | 0.755 |
| 2005 | -0.411 | -0.034 | -0.004 | 0.005 | 0.828 | 0.322 | 1.577 | 0.700 |
| 2006 | -0.433 | -0.047 | -0.352 | -0.271 | 0.112 | 0.206 | -0.020 | -0.815 |
| 2007 | -0.053 | 0.251 | -0.089 | -0.198 | -0.481 | -0.577 | -0.196 | -0.780 |
| 2008 | -0.887 | -0.151 | 0.016 | -0.056 | -0.761 | -0.089 | 0.259 | -1.170 |
| 2009 | -0.416 | -0.354 | -0.188 | 0.289 | -0.243 | -0.485 | 0.294 | -0.067 |
| 2010 | -0.292 | -0.145 | -0.239 | -0.161 | 0.234 | -0.253 | -0.150 | -0.036 |
| 2011 | 0.319 | -0.039 | -0.341 | -0.171 | 0.319 | -0.024 | -0.699 | -0.275 |
| 2012 | 0.348 | 0.452 | 0.192 | -0.099 | 0.085 | 0.202 | -0.054 | 0.710 |
| 2013 | 1.857 | 0.326 | 0.333 | -0.018 | 0.137 | -0.407 | 0.180 | 0.150 |
| 2014 | -0.630 | 0.149 | 0.391 | -0.119 | 0.246 | 0.675 | -0.488 | -1.296 |
| 2015 | -0.948 | 0.072 | -0.092 | -0.300 | 0.616 | -0.312 | -0.718 | 0.347 |
| 2016 | 0.027 | -0.077 | -0.087 | 0.142 | -0.003 | 0.036 | 0.214 | -0.013 |
| 2017 | -0.151 | 0.228 | 0.219 | -0.158 | 0.406 | 0.070 | -0.116 | 0.921 |
| 2018 | -0.295 | -0.158 | -0.221 | 0.094 | -0.031 | 0.117 | -0.268 | -0.511 |
| 2019 | -0.387 | -0.026 | -0.156 | -0.006 | -0.498 | 0.273 | -0.561 | 0.225 |
| 2020 | 0.340 | -0.115 | -0.086 | -0.053 | -0.164 | -0.038 | -0.022 | 0.131 |
| 2021 | 2.152 | 0.657 | 0.259 | 0.189 | 0.148 | 0.144 | 0.717 | 1.027 |
| 2022 | 0.530 | 0.073 | 0.298 | -0.038 | 0.111 | 0.134 | -0.302 | -0.096 |

**Table B2.** Residual values ($US) for the GAM fit on each city by year.

| Year | ATL | AUS | COL | JAX | NAS | OKC | POR | SEA |
|---|---|---|---|---|---|---|---|---|
| 2001 | 0.068 | -0.147 | 0.037 | -0.048 | -0.296 | 0.159 | -0.006 | 0.184 |
| 2002 | -0.793 | -0.530 | -0.232 | 0.119 | -0.463 | -0.369 | -0.052 | -0.686 |
| 2003 | 0.246 | -0.321 | 0.141 | 0.116 | -0.559 | -0.220 | -0.295 | 0.935 |
| 2004 | -0.191 | 0.195 | -0.126 | 0.601 | -0.304 | 0.000 | 1.535 | 0.945 |
| 2005 | -0.411 | 0.195 | -0.186 | -0.203 | 0.857 | 0.530 | 1.807 | 0.884 |
| 2006 | -0.433 | -0.019 | -0.591 | -0.293 | 0.045 | -0.481 | -0.784 | -0.704 |
| 2007 | -0.053 | 0.057 | -0.454 | -0.867 | -0.606 | -0.317 | -0.926 | -1.442 |
| 2008 | -0.887 | -0.190 | -0.397 | -0.025 | -0.967 | -0.607 | -0.414 | -1.151 |
| 2009 | -0.416 | -0.442 | -0.035 | 0.345 | -0.951 | -0.430 | -0.378 | -0.255 |
| 2010 | -0.292 | -0.404 | -0.195 | -0.034 | -0.164 | -0.536 | 0.167 | -0.122 |
| 2011 | 0.319 | -0.103 | -0.149 | 0.145 | 0.026 | -0.513 | -1.677 | -0.200 |
| 2012 | 0.348 | 0.443 | 0.392 | 0.015 | -0.324 | 0.654 | 0.363 | 0.707 |
| 2013 | 1.857 | 0.081 | 0.461 | 0.092 | 0.032 | -0.606 | 0.299 | 0.463 |
| 2014 | -0.630 | 0.172 | 0.516 | -0.078 | 0.688 | 0.971 | -0.090 | -1.100 |
| 2015 | -0.948 | -0.063 | 0.094 | -0.279 | 0.670 | -0.017 | 0.061 | 0.452 |
| 2016 | 0.027 | -0.220 | -0.370 | 0.329 | -0.141 | -0.481 | 0.483 | -0.051 |
| 2017 | -0.151 | -0.066 | 0.200 | -0.200 | 0.331 | -0.061 | -0.271 | 0.853 |
| 2018 | -0.295 | -0.450 | -0.053 | 0.009 | -0.348 | -0.383 | -0.289 | -0.469 |
| 2019 | -0.387 | 0.130 | -0.015 | -0.127 | -0.073 | 0.898 | -0.006 | 0.475 |
| 2020 | 0.340 | 0.036 | -0.113 | -0.051 | -0.308 | 0.032 | -0.385 | -0.006 |
| 2021 | 2.152 | 1.406 | 0.558 | 0.509 | 1.370 | 0.217 | 1.952 | 0.949 |
| 2022 | 0.530 | 0.240 | 0.517 | -0.044 | 1.483 | 1.560 | -1.092 | -0.660 |